\documentclass[letterpaper, 10 pt, conference]{ieeeconf}
\IEEEoverridecommandlockouts     
\usepackage[noadjust]{cite}
\usepackage{graphicx,xcolor,epstopdf}
\usepackage{amsmath,amsfonts,amssymb,mathtools}
\usepackage{subcaption}
\usepackage{tikz,tikz-3dplot,tikzinclude,url,array}
\usepackage{MultiRotorTikz}

\usepackage{hyperref}
\hypersetup{
	colorlinks,
	linkcolor={violet!50!black},
	citecolor={purple!50!black},
	urlcolor={blue!50!black}
}

\tikzset{BlockDiagram/.style={
	block/.style = {draw, rectangle, rounded corners, text centered, minimum height=1cm, minimum width=1cm, text width=2.5cm},
	sum/.style = {draw, circle, thick, minimum size=5mm, node distance=5mm, inner sep=0mm},
	input/.style = {coordinate},
	output/.style = {coordinate},
}} 

\tikzset{BlockDiagram2/.style={
    block/.style = {draw, rectangle, rounded corners, text centered, minimum height=1cm, minimum width=1cm},
    gain/.style = {draw, isosceles triangle, text centered, minimum height=1cm, minimum width=1cm},
    sum/.style = {draw, circle, thick, minimum size=4mm, node distance=5mm, inner sep=0mm},
    input/.style = {coordinate},
    point/.style = {coordinate},
    output/.style = {coordinate},
}} 

\tikzset{FlowChart/.style={
		startstop/.style = {draw,rectangle, rounded corners, fill=green!20,
			minimum width=2cm, minimum height=1cm,
			align=center},
		process/.style = {draw,rectangle, fill=blue!30,rounded corners,
			minimum width=3cm, minimum height=1cm,
			align=center},
		decision/.style = {draw,diamond, aspect=2, fill=green!30,
			minimum width=3cm, minimum height=1cm,
		    align=center},
		io/.style = {draw,trapezium, trapezium stretches body, rounded corners, fill=red!30,
			trapezium left angle=70, trapezium right angle=110,
			minimum width=3cm, minimum height=1cm,
			align=center},
		point/.style = {coordinate},
		arrow/.style = {thick,->},
		sum/.style = {draw, circle, thick, minimum size=5mm, node distance=5mm, inner sep=0mm},
	}
}

\usepackage{amsmath}
\usepackage{amsfonts}
\usepackage{amssymb}

\newtheorem{theorem}{Theorem}

\newtheorem{definition}{Definition}

\def\centerarc[#1](#2)(#3:#4:#5) { \draw[#1] ($(#2)+({#5*cos(#3)},{#5*sin(#3)})$) arc (#3:#4:#5); }

\renewcommand{\vec}{\boldsymbol}

\newcommand{\FI}{\mathcal{I}}
\newcommand{\FB}{\mathcal{B}}
\newcommand{\FR}{\Re}
\newcommand{\FF}{\mathbb{F}}

\newcommand{\RBI}{R_{\FB}^{\FI}}
\newcommand{\RBF}[1]{R_{\FB}^{#1}}

\newcommand{\R}[1]{\mathbb{R}^{#1}}
\newcommand{\I}[1]{\mathbb{I}_{#1}}

\newcommand{\A}{\mathcal{A}}
\newcommand{\B}{\mathcal{B}}
\newcommand{\N}{\mathcal{N}}
\newcommand{\U}{\mathcal{U}}
\newcommand{\F}{\mathcal{F}}
\newcommand{\T}{\mathcal{T}} 
\title{\LARGE \bf Reconfigurable Control of a Class of Multicopters}

\author{Hafiz Zeeshan Iqbal Khan$^{1}$, Jahanzeb Rajput$^{2}$ and Jamshed Riaz$^{3}$
\thanks{$^{1}$Hafiz Zeeshan Iqbal Khan is MS student at Department of Aeronautics \& Astronautics, Institute of Space Technology, Islamabad, and works at National Engineering and Scientific Commission, Islamabad, 44000 Pakistan.}%
\thanks{$^{1}$Jahanzeb Rajput works at National Engineering and Scientific Commission, Islamabad, 44000 Pakistan.}%
\thanks{$^{2}$Jamshed Riaz is Professor at Department of Aeronautics \& Astronautics, Institute of Space Technology, Islamabad, 44000 Pakistan.}%
}

\begin{document}

\maketitle
\thispagestyle{empty}
\pagestyle{empty}

\begin{abstract}
In this paper, a reconfigurable control scheme for a generalized class of multicopters is presented to overcome single or multiple rotor failures. In a multicopter, the rotor failure heavily affects its dynamics and thus its controllability. Limited controllability leads to limited or no reconfigurability. \emph{Available Control Authority Index} has been recently developed \cite{Du2015} as a measure of controllability of multicopters. In this work, the notion of \emph{Available Reduced-Control Authority Index} (ArCAI) is introduced, which shows that in some uncontrollable failures it is still possible to control reduced set of states. Based on this notion, a reconfigurable control scheme is presented, which comprises a \emph{Nonlinear Dynamic Inversion} based baseline control law, a constrained control allocation scheme along with some modifications to incorporate reconfiguration, and a simplified fault detection and isolation technique. Simulation results for a commonly used configuration of Hexacopter are presented in the presence of single and multiple rotor failures.
\end{abstract}

\section{Introduction}
A multicopter can be defined as a flying vehicle in which lift is mainly generated by two or more rotors, e.g., quadcopter, hexacopter, etc. These multicopters are becoming very common in daily life applications due to their ease of construction and control, and their hovering and VTOL capabilities, in contrast to their fixed-wing counterparts. 
As the applications of multicopters are increasing, their performance requirements and safety constraints are getting more stringent, which not only requires better designs but also better and advanced control strategies with embedded safety constraints such as fault tolerance, collision avoidance, etc.

Since rotors are prone to failure and damage, therefore some fault tolerance is usually desirable. 
The faults can generally be classified into three types: actuator faults, sensor faults, and component faults \cite{Hwang2010}. In this work, only actuator faults are considered, specifically failure of individual rotors. In fault-tolerant control, especially subject to actuator failures, a reconfiguration scheme is usually employed. 
Moreover, since multicopters have generally more actuators (rotors) than the number of variables to be controlled, therefore, some control allocation scheme is required to distribute control commands among actuators \cite{Durham2017,Oppenheimer2011}. 

The first task in a fault-tolerant control scheme is \emph{Fault Detection \& Isolation} (FDI). There are many methods available in the literature for fault detection \cite{Hwang2010,Marzat2012}. 
In practical applications, this is usually done by computing residuals, which, in essence, are the difference between estimated and measured signals. In this work, a simplified FDI method based on residuals of individual rotor thrusts is used. It is assumed that individual rotor thrusts are being estimated using some extra sensors e.g. current sensor \cite{Du2018}.

Fault tolerant control of multicopters is an active area of research \cite{Lanzon2014,Mueller2014}. However, failure of a rotor greatly reduces the controllability of a multicopter \cite{Du2015}, which in turn puts severe limitations on the design of the fault-tolerant control system.  In \cite{Du2015}, the notion of \emph{Available Control Authority Index} (ACAI) based on positive controllability \cite{Brammer1972} was introduced, to assess the controllability of multicopters in case of rotor degradation and failures.

The major contribution of this paper lies in the extension of the controllability concept for uncontrollable failures; the basic idea is to evaluate the controllability of a subset of states when the complete system is uncontrollable. The notion of \emph{Available Reduced-Control Authority Index} (ArCAI) is introduced, which shows that even in some uncontrollable failures, it is possible to complete a go-home recovery flight by controlling a reduced set of states. Reconfiguration is done using a constrained control allocation algorithm to compensate for rotor failures. A \emph{Nonlinear Dynamic Inversion} (NDI) based control law \cite{Khan2020} is used with a constrained control allocation scheme, with some modifications to incorporate reconfiguration in the presence of both controllable and uncontrollable rotor failures. Simulation results are presented to show the efficacy of the proposed scheme.


\section{Multicopter Model}
This work focuses on a generalized class of multicopters namely `co-planar multicopters' \cite{Khan2020}, which are defined as follows:

\begin{definition}
A co-planar multicopter is a multicopter in which thrust vectors of all rotors are parallel in Hover for all control inputs.
\end{definition}

It should be noted that almost all common multicopters, e.g., Quadcopter, Hexacopter, fall in the category of these co-planar multicopters. The only exceptions are a few novel designs.
Dynamic model of co-planar multicotpers \cite{Khan2020} can be briefly summarized as follows:

\begin{equation} \label{eq:EOMs}
\begin{split}
\dot{\vec{x}} &= \vec{v}, \\
\dot{\vec{\Phi}} &= \Gamma_1\vec{\omega} + \Gamma_2, \\
\dot{\vec{v}} &= ge^{\FI}_3 - \frac{F_T}{m}\RBI e^{\FB}_3 - \frac{\kappa_{D}}{m}\vec{v}\|\vec{v}\|, \\
\dot{\vec{\omega}} &= J^{-1}\left(\vec{T} - \vec{\omega}\times J \vec{\omega} - \kappa_{R} \vec{\omega}\right)
\end{split}
\end{equation}
where $\vec{x}$ and $\vec{v}$ are multicopter's position and velocity in inertial reference frame, respectively, $\vec{\Phi} = [\phi,\theta,\psi]^T$ are Euler angles, $\vec{\omega} = [p,q,r]^T$ are body angluar rates, $\vec{T} = [L,M,N]^T$ are roll, pitch and yaw moments, respectively, $F_T$ is total thrust, $g$ is acceleration due to gravity, $m$ is mass of multicopter, $J=\mathrm{diag}(J_x,J_y,J_z)$ is inertia matrix, $\kappa_D$ is drag factor, $\kappa_R$ is rotational damping factor \cite{Mueller2014,Faessler2018,Kai2017},  $e_i^\FF$ are unit vectors in $\FF$ frame, $\RBI$ is rotation matrix from body frame ($\FB$) to inertial frame ($\FI$), $\Gamma_1 = \mathrm{diag}(1,\cos\phi,\cos\phi/\cos\theta)$ is invertible $\forall\; \phi,\theta \in (-\pi/2,\pi/2)$, and
\[\Gamma_2 = \begin{bmatrix}
q\sin\phi\tan\theta + r\cos\phi\tan\theta \\ -r\sin\phi \\ q\sin\phi/\cos\theta \end{bmatrix}\]

\begin{figure}
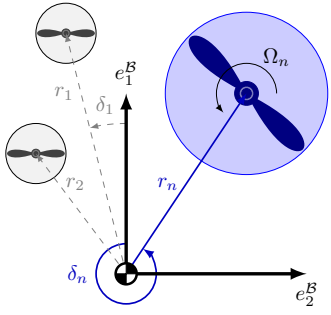

	\centering \scalebox{0.8}{ \includetikzgraphics[MultipleRotorPlanar]{TikzDatabase.tex}}
	\caption{A generalized co-planar multicopter schematic (Top View)}
	\label{fig:rotorschem}
\end{figure}

\begin{figure}
	\centering
    \begin{subfigure}[b]{0.23\textwidth}
    \centering
    \def\Seq{{1,2,1,2,1,2}}
	\begin{tikzpicture}
	\node [NRotor,minimum width=0.8\linewidth, N=6, RotationSeq = \Seq, IsDucted = 1, IsDir = 1, IsNumbered = 1] at (0,0) {};
	\end{tikzpicture}
    \caption{PNPNPN}
    \end{subfigure}
    \begin{subfigure}[b]{0.23\textwidth}
    \centering
    \def\Seq{{1,1,2,2,1,2}}
	\begin{tikzpicture}
	\node [NRotor,minimum width=0.8\linewidth, N=6, RotationSeq = \Seq, IsDucted = 1, IsDir = 1,IsNumbered = 1] at (0,0) {};
	\end{tikzpicture}
    \caption{PPNNPN}\label{fig:HexConf}
    \end{subfigure}
	\caption{Typical \emph{co-planar} multicopter configurations}
	\label{fig:PlanarMRCs}
\end{figure}
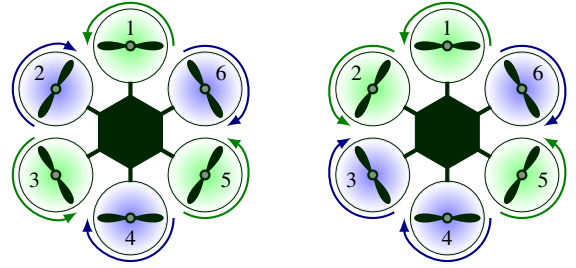

Actuator dynamics is modelled as first order lag.
\begin{equation}\label{eq:motor}
  \frac{\Omega_n(s)}{\Omega_{cmd-n}(s)} = \frac{1}{\tau_{motor}s + 1}, \qquad \forall\,n \in [1,\N]
\end{equation}
where $\Omega_{cmd-n}$ is commanded RPM and $\Omega_{n}$ is actual RPM of $n$th rotor. Considering the schematic shown in Fig. \ref{fig:rotorschem}, the total thrust and moments of all $\N$ rotors can be computed as follows,
\begin{equation}\label{eq:EffModel}
  \tau = \begin{bmatrix} F_T \\ \vec{T} \end{bmatrix} = \sum_{n=1}^{\N} \begin{bmatrix} \epsilon_n \\ \epsilon_n r_n \sin\delta_n \\ \epsilon_n r_n \cos\delta_n \\ \epsilon_n \gamma_n \kappa_\mu \end{bmatrix} f_n \stackrel{\vartriangle}{=} B \vec{f}
\end{equation}
where $\vec{f}=[f_1,\cdots,f_\N]^T$, $B$ is control effectiveness matrix, and $f_n = \kappa_T \Omega_n^2$ is thrust of $n$-th rotor, $r_n$ is the distance from CG to its center, $\delta_n$ is its angle from $x$-axis, $\gamma_n$ is direction of rotation ($+1$ for counter-clockwise and $-1$ for clockwise), as shown in Fig. \ref{fig:rotorschem}, and $\kappa_T,\kappa_\mu$ are thrust and torque factors, respectively. Here $\epsilon_n$ is health of $n$th rotor. In normal condition $\epsilon_n = 1,\,\forall n\in[1,\N]$; while in the case of rotor failure, $\epsilon_n = 0$ for all failed rotors. For symmetric configurations $r_n = r$, $\delta_n = (2\pi/\N)(n-1)$ and $f_n \in [0,F_{max}]$ $\forall\,\,n\in [1,\N]$, while rotors are numbered counter-clockwise, the configuration (all $\gamma_n$) is also encoded as string of length $\N$ with `P' for $\gamma = +1$ and `N' for $\gamma = -1$. For example in hexacopters `PNPNPN' configuration represents $\vec{\gamma} = [+1,-1,+1,-1,+1,-1]$ while `PPNNPN' configuration represents $\vec{\gamma} = [+1,+1,-1,-1,+1,-1]$ as shown in Fig. \ref{fig:PlanarMRCs}.

\section{Controllability of Co-Planar Multicopters}
Since in a co-planar multicopter maximum of four variables can be controlled at a time, which are usually either its attitude and altitude or position and heading; therefore, for controllability analysis, attitude and altitude dynamics are considered. Horizontal position dynamics doesn't account for much in controllability analysis since it can be considered as a set of outer-loops in roll and pitch angles. When the nonlinear dynamics of aircraft is linearized about hover condition i.e. $(\phi = 0,\theta = 0)$, we get following linear model.

\begin{equation}\label{eq:LinMod}
  \dot{x} = \A x + \B \underbrace{(\tau-G)}_{u}
\end{equation}
where $x = [h,\phi,\theta,\psi,v_h,p,q,r]^T$ is a state vector of linear model, $\tau$ is net thrust and moment of all rotors as in Eq. \eqref{eq:EffModel}, $h$ is altitude and $v_h$ is climb velocity. Since for each rotor thrust ($f_n\in[0,F_{max}]$), thus $\vec{f}\in \F = \prod_{n=1}^{\mathcal{N}}[0,F_{max}]$ and $\tau \in \T=\{\tau | \tau = B\vec{f}, \vec{f} \in \F\}$, which implies $u \in \U = \{u|u=\tau - G, \tau \in \T\}$, where $G = [mg, 0, 0, 0]^T$, and,

\[\A = \begin{bmatrix}
        O_{4\times4} & \I{4} \\
        O_{4\times4} & O_{4\times4}
      \end{bmatrix}, \qquad\qquad
      \B = \begin{bmatrix}
            O_{4\times4} \\
            J_f^{-1}
          \end{bmatrix}\]
where, $J_f = \mathrm{diag}(-m,J)$, $O_{n\times m}$ is a zero matrix of specified size.

The main restriction in the controllability of a mulitcopter is usually not due to rank deficient controllability gramian but due to non-negative and limited individual rotor thrusts. The \emph{Positive Controllability Theory} was proposed in \cite{Brammer1972} for LTI systems with bounded control inputs and was used for controllability analysis of multicopters in hover by \cite{Du2015}. The following theorem from \cite{Du2015} is restated here without proof for completeness:

\begin{theorem}\label{TH5:1}
The following conditions are necessary and sufficient for the controllability of the system:
\begin{enumerate}
  \item Pair $(\A,\B)$ is \emph{Controllable}. i.e. Controllability matrix is of full rank.
  \item There is no real eigenvector $v$ of $\A^T$ satisfying $v^T\B u \leq 0,\;\forall\;u \in \U$.
\end{enumerate}
\end{theorem}

\subsection{Available Control Authority Index}
Since second condition in Theorem \ref{TH5:1} is quite difficult to test, as it is not possible to check all $u\in\U$, so \cite{Du2015} proposed a measure (ACAI) as defined in \eqref{eq:ACAIdef}, and proved that for system described in Eq. \eqref{eq:LinMod}, this second condition is equivalent to the requirement that the ACAI $\rho > 0$. Therefore, from Theorem \ref{TH5:1}, system \eqref{eq:LinMod} is controllable iff $\mathrm{rank}\,\mathcal{C}(\A,\B) = 8$ and $\rho > 0$.

\begin{definition}
Available Control Authority Index (ACAI) is defined as
\begin{equation}\label{eq:ACAIdef}
  \rho \stackrel{\vartriangle}{=} \begin{cases}
                \min(\|\tau-G\|: G \in \T, \tau \in \partial\T) \\
                -\min(\|\tau-G\|: G \in \T^C, \tau \in \partial\T)
              \end{cases}
\end{equation}
\end{definition}
where $\partial\T$ represents the boundary of $\T$ and $\T^C$ represents the complement of $\T$.

\begin{figure}
	\centering
    \begin{subfigure}[b]{0.235\textwidth}
    \centering
    \includegraphics[width=\linewidth,trim={1.6cm 0 1.3cm 0.7cm},clip] {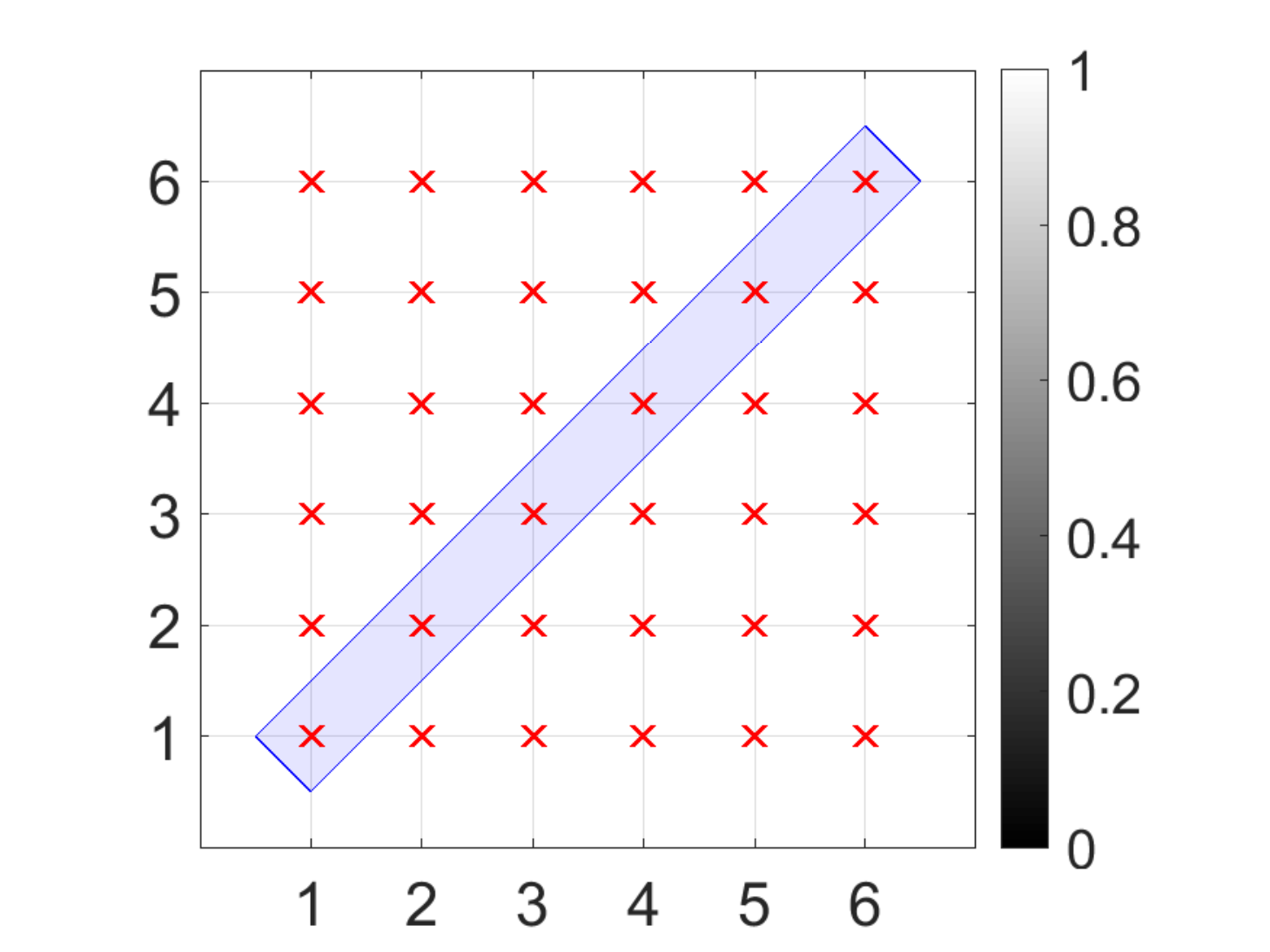}
    \caption{PNPNPN}
    \end{subfigure}
    \begin{subfigure}[b]{0.235\textwidth}
    \centering
    \includegraphics[width=\linewidth,trim={1.6cm 0 1.3cm 0.7cm},clip] {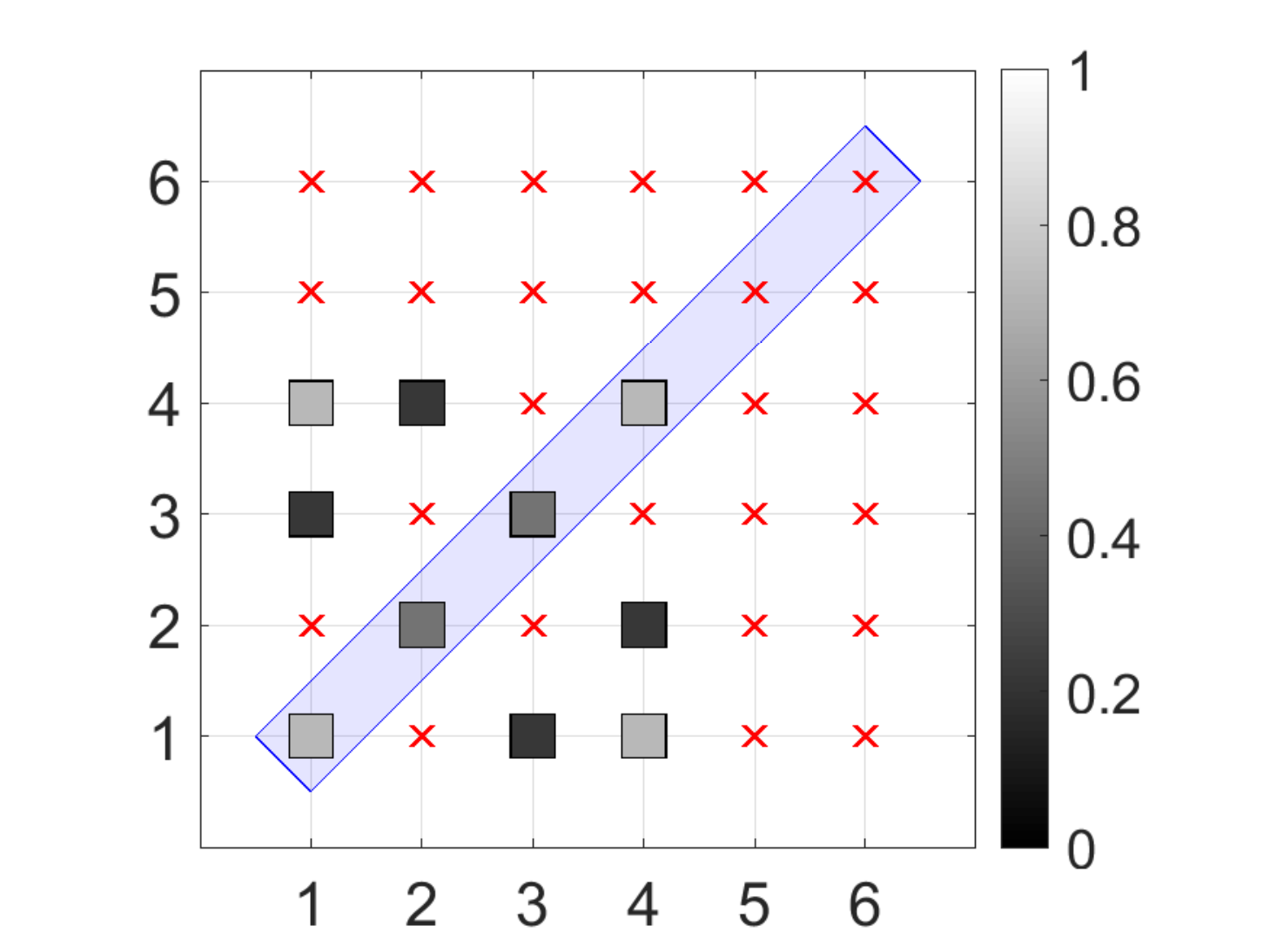}
    \caption{PPNNPN}
    \end{subfigure}
	\caption{ACAI of Two Common Hexacopter Configurations with up to two rotor failures}
	\label{fig:HexACAI}
\end{figure}

Fig. \ref{fig:HexACAI} shows the ACAI of two common Hexacopter configurations ('PNPNPN' and 'PPNNPN'), up to two rotor failures. Each $ij$th grid point corresponds to $i$ and $j$ rotor failures, entries on diagonal show single rotor failure, `$\times$' means uncontrollable failures ($\rho \leq 0$), the failures which lead to uncontrollable systems, while shading of the boxes shows how much controllable those failures are, i.e. ACAI. It can be seen from Fig. \ref{fig:HexACAI} that `PNPNPN' is uncontrollable for any rotor failure, while `PPNNPN' configuration can remain controllable for the failure of rotors 1 to 4,  but uncontrollable if either rotor 5 or 6 fails.

\subsection{Available Reduced Control Authority Index}
As seen from the last section, there are quite a few cases in which failure results in uncontrollable system, which puts severe limitation on fault tolerance control. Therefore, though all states are not controllable, it would be of great importance if a critical subset of states can be controlled. This can protect multicopter from crashing, and make it possible to land multicopter safely in such critical scenarios. Considering the linear system \eqref{eq:LinMod}, if we assume $J$ is diagonal, which it usually is, then $J_f^{-1}$ will also be diagonal; this leads to four decoupled systems as follows,
\[\ddot{h} = -F_T/m , \quad \ddot{\phi} = L/J_x, \quad \ddot{\theta} = M/J_y, \quad \ddot{\psi} = N/J_z\]

Let us define a set of reduced linear models, by considering only three of above mentioned decoupled systems at a time. which leads to the following four systems,

\begin{equation}\label{eq:LinModred}
  \dot{x}_{\alpha} = \underbrace{\begin{bmatrix}
        O_{3\times3} & \I{3} \\
        O_{3\times3} & O_{3\times3}
      \end{bmatrix}}_{\A_\alpha} x_{\alpha} + \underbrace{\begin{bmatrix}
        O_{3\times3} \\
        J_{\alpha}^{-1}
      \end{bmatrix}}_{\B_\alpha}(\tau_\alpha - G_\alpha)
\end{equation}
where $\alpha = h,\,\phi,\,\theta\,\text{or}\,\psi$, and $x_{\alpha}$ contains all states except corresponding state and its derivative, e.g.  $x_\psi = [h,\phi,\theta,v_h,p,q]^T$. Similarly, $J_\alpha,\tau_\alpha,G_\alpha$ are $J,\tau,G$ with corresponding row and/or column removed, respectively, e.g $J_h = J$, $G_h = [0,0,0]^T$, $G_\psi=[mg,0,0]^T$, $\tau_h = [L,M,N]^T$, $\tau_\psi = [F_T,L,M]^T$ etc.

\begin{definition}
Available Reduced Control Authority Index (ArCAI) is defined as
\begin{equation}\label{eq:ArCAIdef}
  \rho_r \stackrel{\vartriangle}{=} \begin{cases}
                \min(\|\tau_\alpha-G_\alpha\|: G_\alpha \in \T_\alpha, \tau_\alpha \in \partial\T_\alpha) \\
                -\min(\|\tau_\alpha-G_\alpha\|: G_\alpha \in \T_\alpha^C, \tau_\alpha \in \partial\T_\alpha)
              \end{cases}
\end{equation}
\end{definition}

Therefore, from Theorem \ref{TH5:1}, reduced systems (\ref{eq:LinModred}) are controllable if $\mathrm{rank}\, \mathcal{C}(\A_\alpha,\B_\alpha) = 6$ and $\rho_r > 0$. The same algorithm that was presented by \cite{Du2015} to compute ACAI efficiently is used to compute ArCAI with minor modifications.

\begin{figure}
	\centering
    \begin{subfigure}[b]{0.235\textwidth}
    \centering
    \includegraphics[width=\linewidth,trim={1.6cm 0 1.3cm 0.7cm},clip] {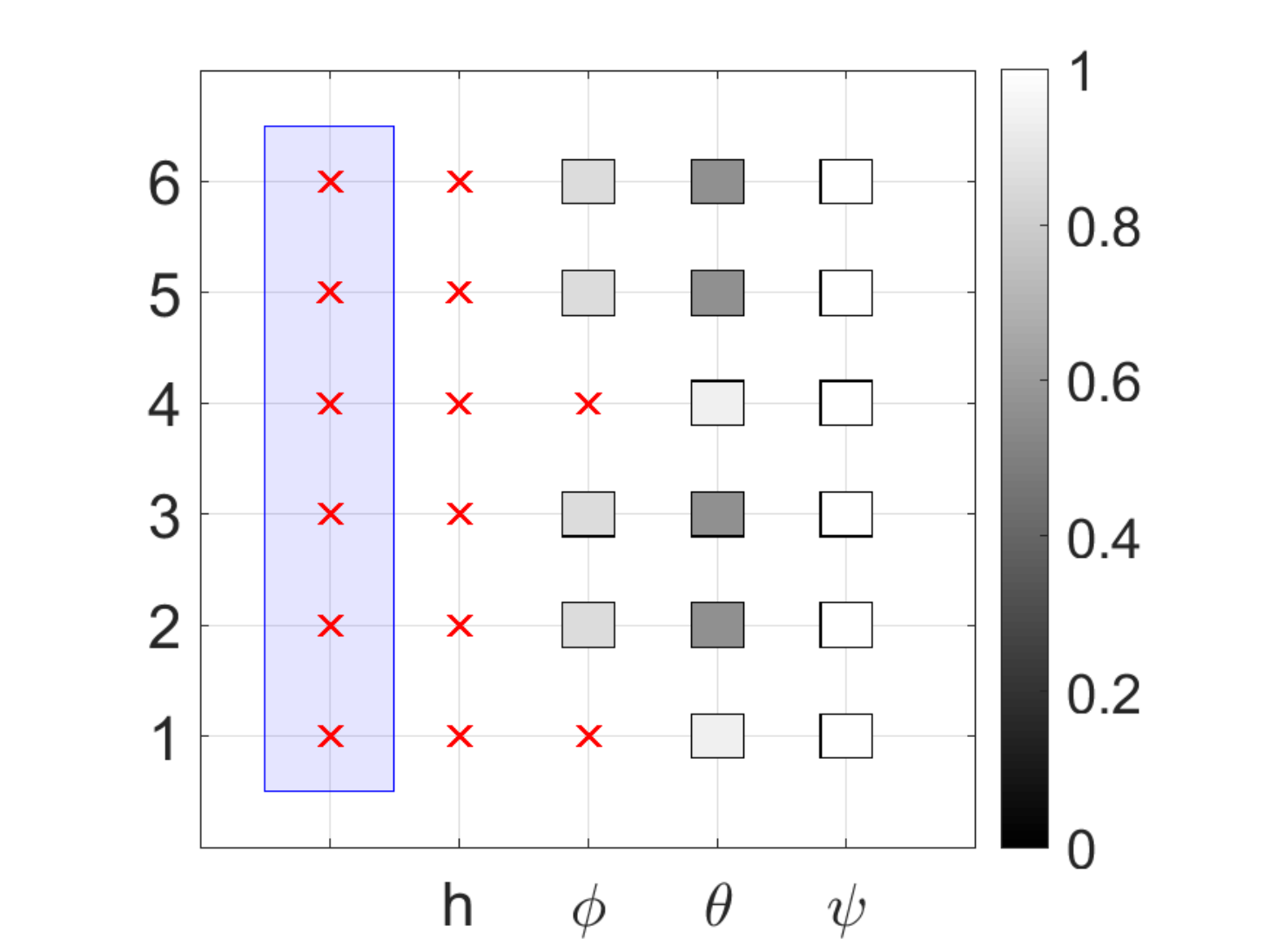}
    \caption{PNPNPN}
    \end{subfigure}
    \begin{subfigure}[b]{0.235\textwidth}
    \centering
    \includegraphics[width=\linewidth,trim={1.6cm 0 1.3cm 0.7cm},clip] {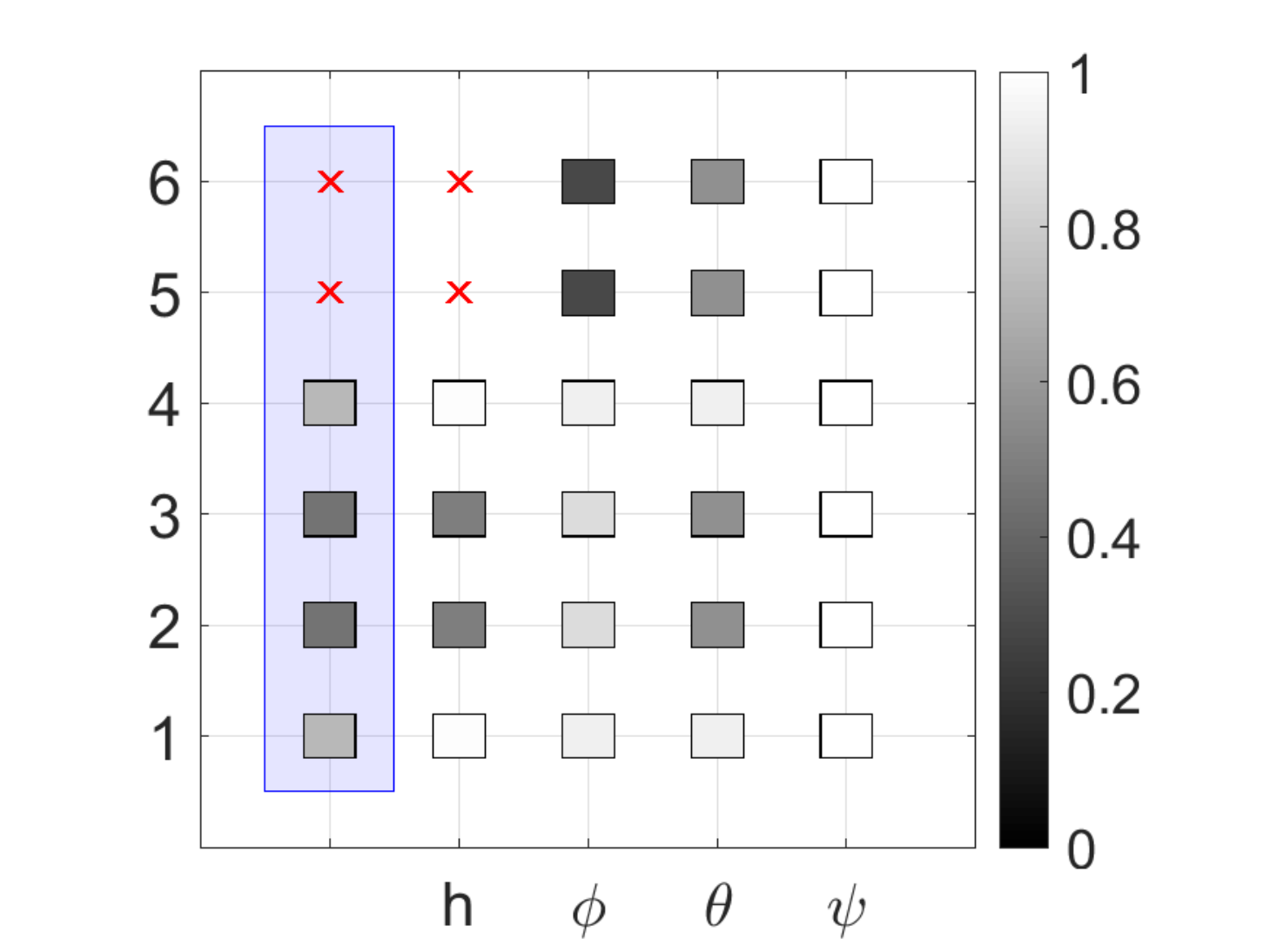}
    \caption{PPNNPN}
    \end{subfigure}
	\caption{ArCAI of Two Common Hexacopter Configurations with single rotor failures}
	\label{fig:HexArCAI}
\end{figure}

Fig. \ref{fig:HexArCAI} shows ArCAI of `PNPNPN' and `PPNNPN' hexacopters. First column shows full state controllability (ACAI), while remaining three columns show reduced controllability (ArCAI) with one state, labeled along horizontal axis, left uncontrolled. It can be seen that, even though for `PNPNPN' any rotor failure is not fully controllable, yet if we choose not to control $\theta$ or $\psi$, the remaining states become controllable in hover for any rotor failure. Similarly if we let $\phi$ to be uncontrolled, in case of failures in rotor 2, 3, 5 and 6, the reduced system becomes controllable. Also as we know for `PPNNPN' configuration failures in rotor 5 or 6 are not controllable but if we choose not to control any of the attitude angles, then the remaining states become controllable. Here it must be noted that ArCAI is defined only for hover case, so it doesn't give any insight about non-hover conditions. This limitation of analysis becomes very significant especially if we let either $\phi$ or $\theta$ uncontrolled, because with that multicopter goes out of hover conditions almost instantaneously, which may cause instability. But this analysis shows that even in some `uncontrollable' failures we can let $\psi$ be uncontrolled, and then it is possible to safely land the spinning multicopter.

\section{Reconfigurable Control Design}
The main idea behind this scheme is to divide reconfigurable control scheme intro three parts, the baseline control law which in this case is a robust nonlinear dynamic inversion based control, fault detection and isolation algorithm, and most importantly the constrained control allocation, using which, fault tolerance in the presence of both controllable and uncontrollable cases is achieved.

\subsection{NDI Controller}
\begin{figure*}
	\centering
    \includegraphics[width=0.7\linewidth]{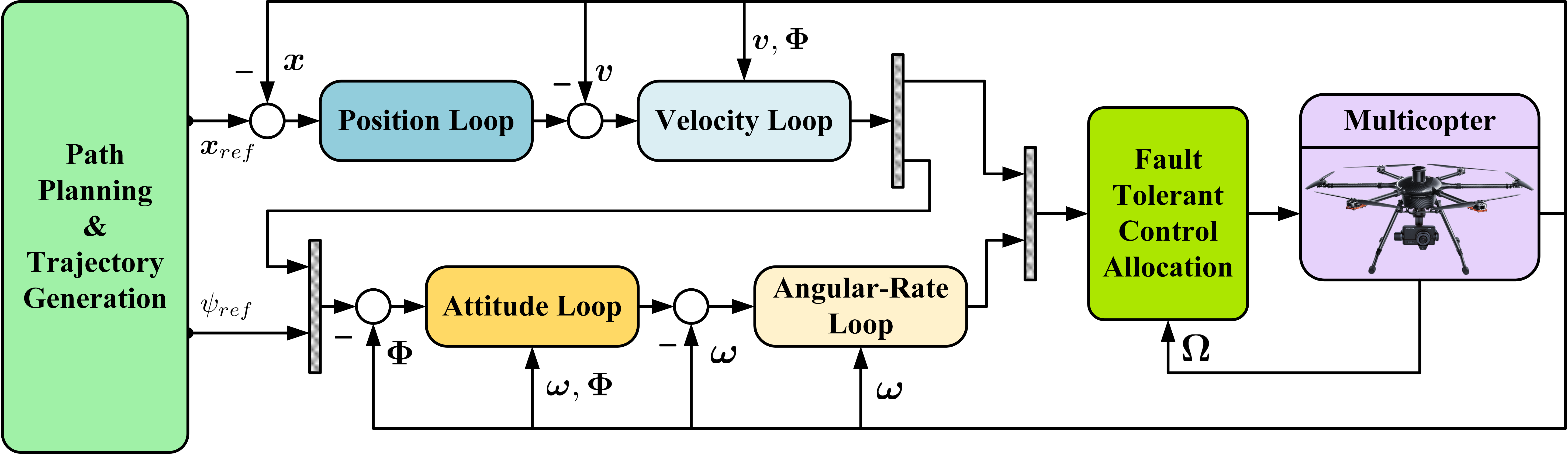}
	\caption{Complete Control Structure}
	\label{fig:CStruct}
\end{figure*}

The successive loop approach of an NDI controller presented in \cite{Khan2020} was used. 
Time-scale separation between successive loops is assumed and can be ensured by appropriate selection of gains in each loop. It was shown in \cite{Khan2020} that the controller \eqref{eq:NDICont}, although dependent on system parameters, ensures robust stability (\emph{uniform ultimate boundedness}) in the presence of unknown but bounded uncertainties in system parameters.

Let $\vec{e}_\omega = \vec{\omega}_{d} - \vec{\omega}$, $\vec{e}_\Phi = \vec{\Phi}_{d} - \vec{\Phi}$, $\vec{e}_v = \vec{v}_{d} - \vec{v} = [e_u,e_v,e_w]^T$, and
$\vec{e}_x = \vec{x}_{d} - \vec{x}$. To maintain time-scale separation between successive loops it was assumed in \cite{Khan2020} that the references ($\vec{\omega}_{d},\vec{\Phi}_{d},\vec{v}_{d},\vec{x}_{d}$) are sufficiently slow, such that $\dot{\vec{\omega}}_{d}  \approx \vec{0}$, $\dot{\vec{\Phi}}_{d} \approx \vec{0}$, $\dot{\vec{v}}_{d} \approx \vec{0}$, and $\dot{\vec{x}}_{d} \approx 0$. This leads to following control law,
\begin{equation} \label{eq:NDICont}
\begin{split}
 \vec{T}_d &\stackrel{\vartriangle}{=} \vec{\omega}\times\tilde{J}\vec{\omega} + \tilde{\kappa}_{R} \vec{\omega} + \tilde{J} K_\omega \vec{e}_\omega, \\
\vec{\omega}_{d} &\stackrel{\vartriangle}{=} \Gamma_1^{-1}\left(K_\Phi \vec{e}_\Phi - \Gamma_2 \right), \\
  \begin{bmatrix} \tan\phi_{d} \\ \sin\theta_{d}
\end{bmatrix} & \stackrel{\vartriangle}{=} \frac{\tilde{m}}{F_T\cos\phi}\Gamma_3^{-1}\left(\begin{bmatrix} k_u e_u \\k_v e_v \end{bmatrix} + \frac{\tilde{\kappa}_{D}\|\vec{v}\|}{\tilde{m}} \begin{bmatrix} u \\ v \end{bmatrix}\right), \\
F_{T_d} &\stackrel{\vartriangle}{=}  \frac{\tilde{m}}{\cos\phi\cos\theta}\left(g - k_w e_w - \frac{\tilde{\kappa}_{D}\|\vec{v}\|}{\tilde{m}} w\right), \\
\vec{v}_{d} &\stackrel{\vartriangle}{=} K_x \vec{e}_x
\end{split}
\end{equation}
where, $\tilde{J}$, $\tilde{\kappa}_{R}$, $\tilde{m}$ and $\tilde{\kappa}_D$ are known or estimated values for $J$, $\kappa_{R}$, $m$ and $\kappa_{D}$, respectively, and $K_\omega > 0$, $K_\Phi > 0$, $K_v = \mathrm{diag}(k_u,k_v,k_w) > 0$, and $K_x > 0$ are positive definite controller gains matrices and
\[
\Gamma_3(\psi) = \begin{bmatrix} -\sin\psi & -\cos\psi \\ \cos\psi & -\sin\psi \end{bmatrix}.
\]
It can be seen clearly that $\Gamma_3(\psi)$ is invertible for $\psi \in [-\pi,\pi]$. The overall control structure is summarized in Fig. \ref{fig:CStruct}. 
\subsection{Fault Detection \& Isolation}
\begin{figure}
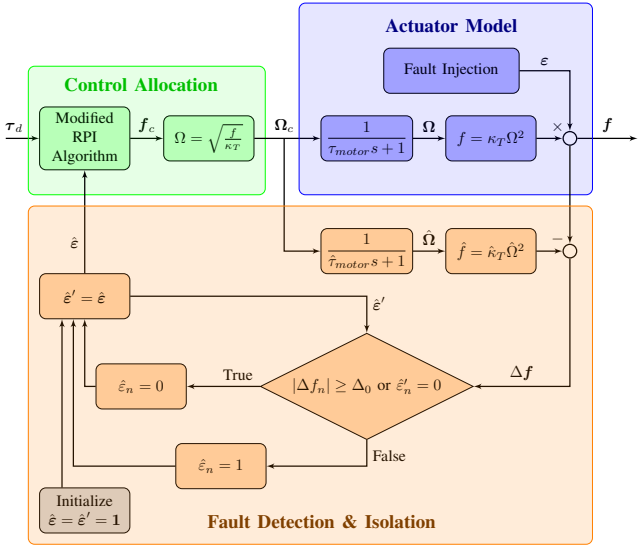

	\centering
	\centering \scalebox{0.6}{\includetikzgraphics[FDIFlow]{TikzDatabase.tex}}
	\caption{Fault Tolerant Control Allocation with a Simplified FDI Algorithm}
	\label{fig:FDIAlgo}
\end{figure}

In this work, a simplified method based on the residuals of individual rotor thrust is used as described in Fig. \ref{fig:FDIAlgo}. The basic idea is to compute residuals ($|\Delta\vec{f}|$) between actual thrust of the rotor, estimated from feedback, and thrust computed from onboard model. This obviously requires the rotor thrusts to be observable using measured signals. This requires some extra sensors, specifically for multicopters with more than four rotors \cite{Du2018}, e.g. current sensor etc., and they are assumed to be present. In nominal scenarios, these residuals remain negligible ($|\Delta\vec{f}| \approx 0$), but in case of occurrence of fault at any rotor, corresponding $|\Delta f_n|$ will increase and when it reaches a certain threshold ($\Delta_0$), then the corresponding rotor is declared as faulty ($\varepsilon_n=0$). The threshold limit ($\Delta_0$) can be considered as a tunable parameter. It is highly dependent on the choice of sampling time, too small a value can cause false detections e.g. there is no fault but algorithm detects one, while too high a value can ignore actual faults. During initial transients, when multicopter starts, the FDI algorithm can also make false detections. To overcome this, FDI algorithm does not start with controller ($t=0$, instead it starts after some time ($t=t_0>0$). Where $t_0$ is selected such that initial large transients are sufficiently decayed to trigger any false detection. Moreover, once an actuator is declared faulty, it is isolated i.e. it is considered faulty for all future times.

\subsection{Control Allocation}

For co-planar multicopters the problem of distribution of control commands to actuators can be written as a linear constrained allocation problem. which can be defined as follows: find the control vector $\vec{f}$, such that
\begin{equation}\label{eq:CAprob}
  \tau_d = B \vec{f}
\end{equation}
where $\vec{f} \in \F$ and $\tau_d$ is desired total forces and moments, and $B$ is control effectiveness matrix \eqref{eq:EffModel}. Eq. \eqref{eq:CAprob} can be written in expanded form as,
\[
\tau_d = \begin{bmatrix}
  \epsilon_1                     & \epsilon_2                     & \cdots & \epsilon_\N  \\
  \epsilon_1 r_1 \sin\delta_1    & \epsilon_2 r_2 \sin\delta_2    & \cdots & \epsilon_\N r_\N \sin\delta_\N \\
  \epsilon_1 r_1 \cos\delta_1    & \epsilon_2 r_2 \cos\delta_2    & \cdots & \epsilon_\N r_\N \cos\delta_\N \\
  \epsilon_1 \gamma_1 \kappa_\mu & \epsilon_2 \gamma_2 \kappa_\mu & \cdots & \epsilon_\N \gamma_\N \kappa_\mu \\
  \end{bmatrix}\begin{bmatrix} f_1 \\ f_2 \\ \vdots \\ f_\N \end{bmatrix}
\]

Once the control allocation algorithm computes $\vec{f}$, actuator input (RPM) can be computed as.
\begin{equation}\label{eq:CArpm}
  \Omega_{cmd-n} = \sqrt{\frac{f_n}{\kappa_T}}, \qquad \forall\,n\in[1,\N]
\end{equation}

In this work \emph{Redistributed Pseudo Inverse} (RPI) method, which is based on weighted \emph{Moore-Penrose} pseudo-inverse, is used to find thrust vector ($\vec{f}$) \cite{Oppenheimer2011,Khan2018}. 
Redistributed Pseudo Inverse (RPI) solution of control allocation problem is given as
\begin{equation}\label{eq:RPI}
  \vec{f} = \vec{c} + W^{-1}B^T\left(B W^{-1}B^T + \varepsilon \I{m}\right)^{-1}\left(\vec{\tau}_d - B_0 \vec{c}\right)
\end{equation}
where $B_0$ is the original effectiveness matrix, $B$ is the modified effectiveness matrix after each iteration if saturation occurs, $W$ is a weighting matrix, $\vec{c}$ is saturation/offset vector, $\vec{d}$ is the corresponding index vector. The $\varepsilon\mathbb{I}$ term is added to avoid singularity in pseudo-inverse \cite{Fossen2011}. To incorporate fault tolerance into the RPI method another vector $\vec{d}$ is introduced and depending upon the estimated rotor health $\hat{\epsilon}$ from FDI algorithm, all elements of $\vec{d}$ corresponding to faulty rotors are set as one and remaining elements as zero. When RPI gets any non-zero element in $\vec{d}$ at any time instant it changes the corresponding column of $B$ and $B_0$ to zero. Once the solution is computed using Eq. \eqref{eq:RPI}, it is checked, if one or more rotors reach saturation, then corresponding elements of vector $\vec{c}$ are set equal to that saturation limit, while the corresponding elements of vector $\vec{d}$ are set equal to one, and corresponding columns of the effectiveness matrix are set equal to zero. When failures that lead to uncontrollable configurations are detected, instead of solving complete control allocation problem, reduced problem ($\tau_\alpha^d = B_\alpha \vec{f}$) is solved,
where $\tau_\alpha^d$ is the desired value of reduced torque vector as defined in \eqref{eq:LinModred}, and $B_\alpha$ is reduced effectiveness matrix (row corresponding to uncontrolled channel is removed). This on-board classification of detected failure being controllable or not, can be easily done by pre-flight offline evaluations of ACAI for all possible failures and also of ArCAI for uncontrollable ones, to plan which states can be left uncontrolled.   

\section{Results \& Discussion}
\begin{figure}
  \centering
  \includegraphics[width=0.8\linewidth,trim={0.5cm 0.5cm 0cm 0.8cm}, clip]{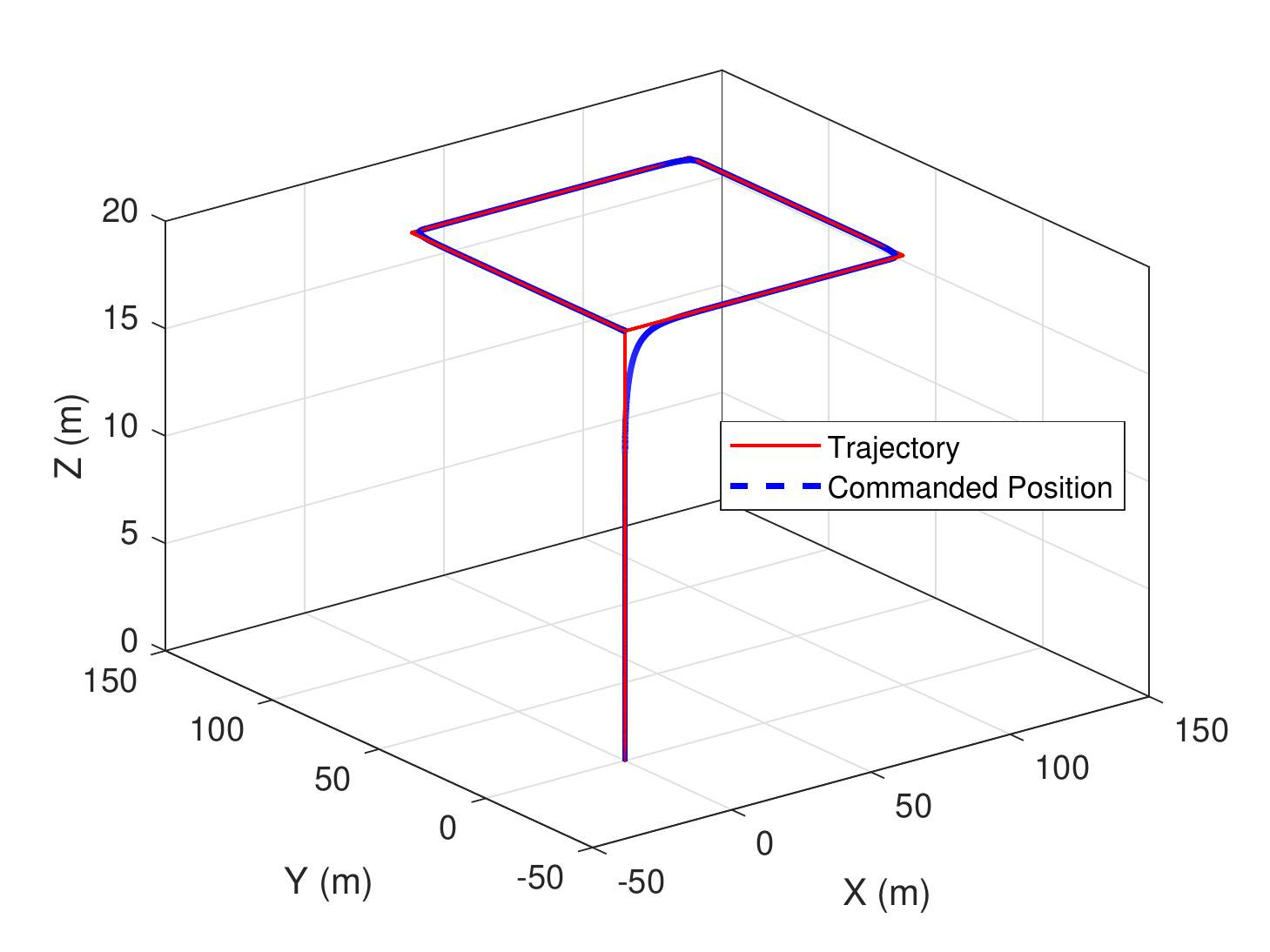}
  \caption{Typical Trajectory Profile}\label{fig:Traj}
\end{figure}

To demonstrate the fault tolerance of proposed scheme in the presence of controllable and uncontrollable faults, different faults are injected in simulation and results are presented in this section. Parameters of a \emph{Hexacopter} (`PPNNPN' configuration) shown in Fig. \ref{fig:HexConf} \cite{Du2015,Khan2020} are used in simulation. For simulation, a typical trajectory shown in Fig. \ref{fig:Traj} is used.

\subsection{Controllable Rotor Failures}
\begin{figure}
  \centering
  \includegraphics[width=0.9\linewidth]{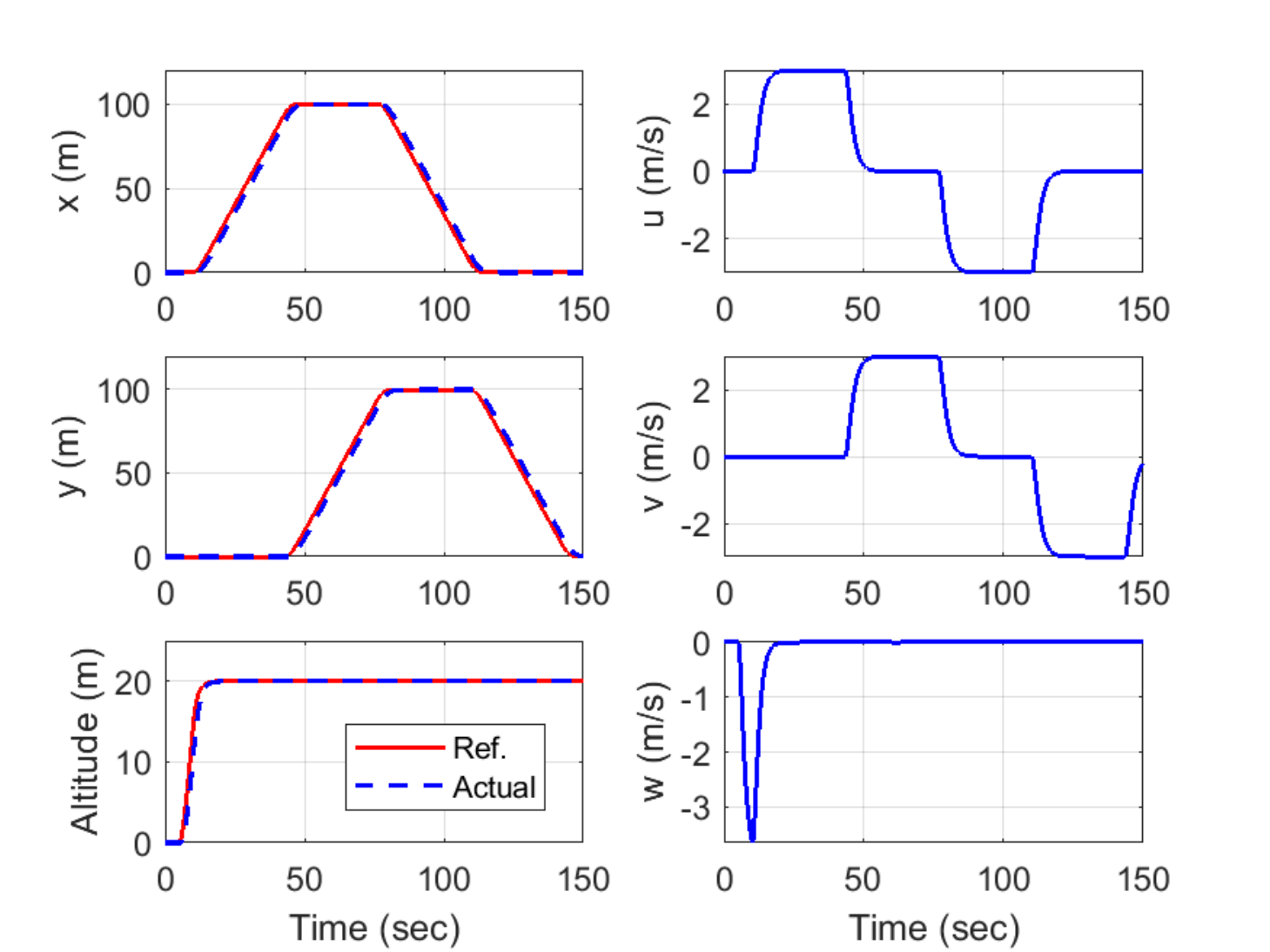}
  \caption{Controllable Faults: Position \& Velocity}\label{fig:FTCPosVel}
\end{figure}

\begin{figure}
  \centering
  \includegraphics[width=0.9\linewidth]{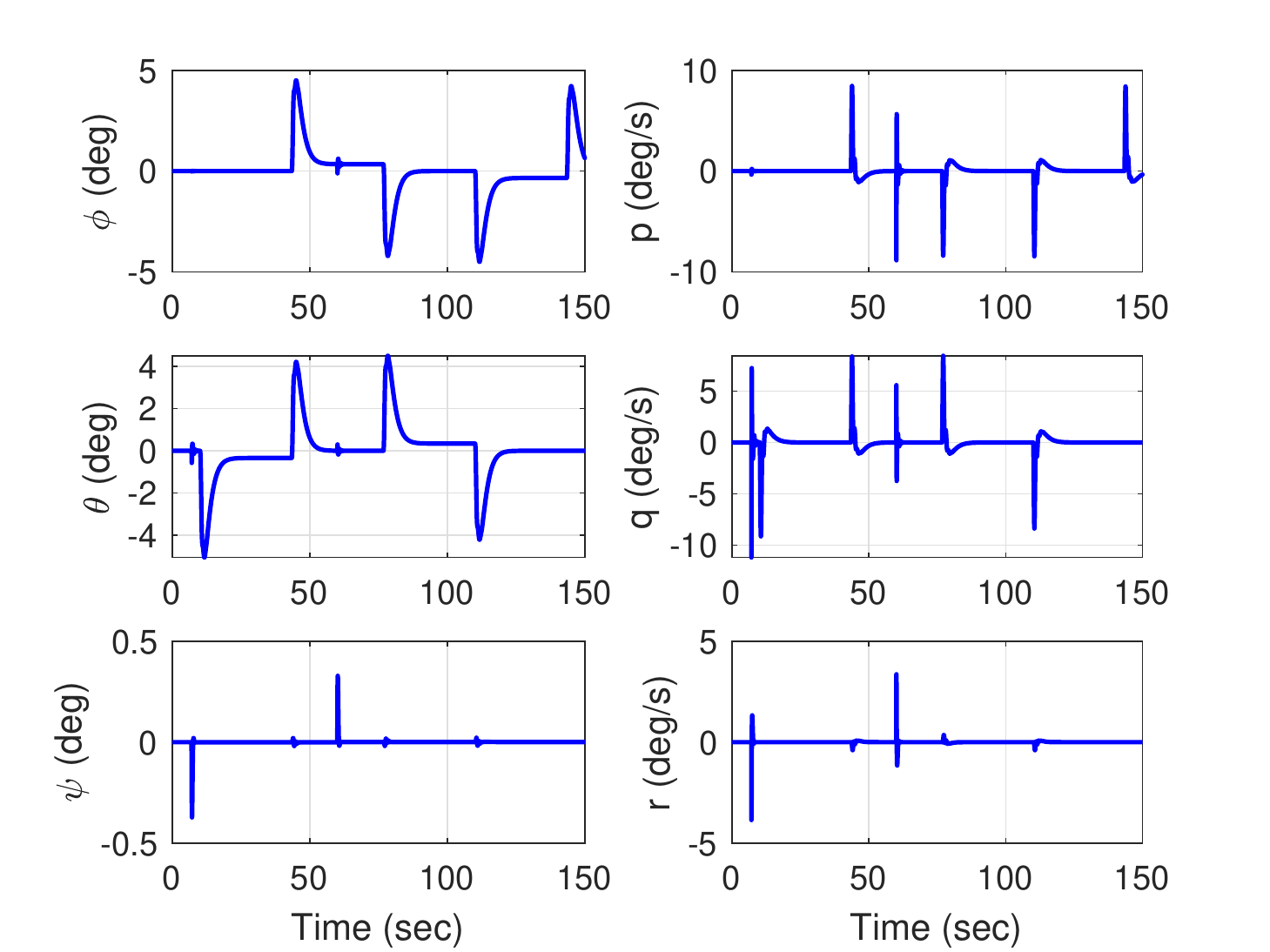}
  \caption{Controllable Faults: Attitude \& Rates}\label{fig:FTCAttitude}
\end{figure}

During climb phase rotor 1 is disabled at 7 seconds, and at 60 seconds rotor 3 fails during forward flight. Fig. \ref{fig:FTCPosVel} shows that despite these failures multicopter continues to track the desired trajectory quite accurately. While in Fig. \ref{fig:FTCAttitude} it can be seen that when failure happens, there is a slight but visible error of a fraction of degree in attitudes, and about 10 deg/s spikes in body angular rates as the failure happens but it gets corrected by the scheme in a short time. 

\subsection{Uncontrollable Rotor Failures}
As shown in Fig. \ref{fig:HexACAI}, in `PPNNPN' type Hexacopter, rotor 5 or 6 failures are uncontrollable. At 60 seconds during forward flight phase, rotor 5 is failed intentionally and control scheme is reconfigured such that it lets yaw ($\psi$) be uncontrolled. From previous analysis we know that reduced controllability is available (Fig. \ref{fig:HexArCAI}), and results confirm it.

\begin{figure}
  \centering
  \includegraphics[width=0.9\linewidth]{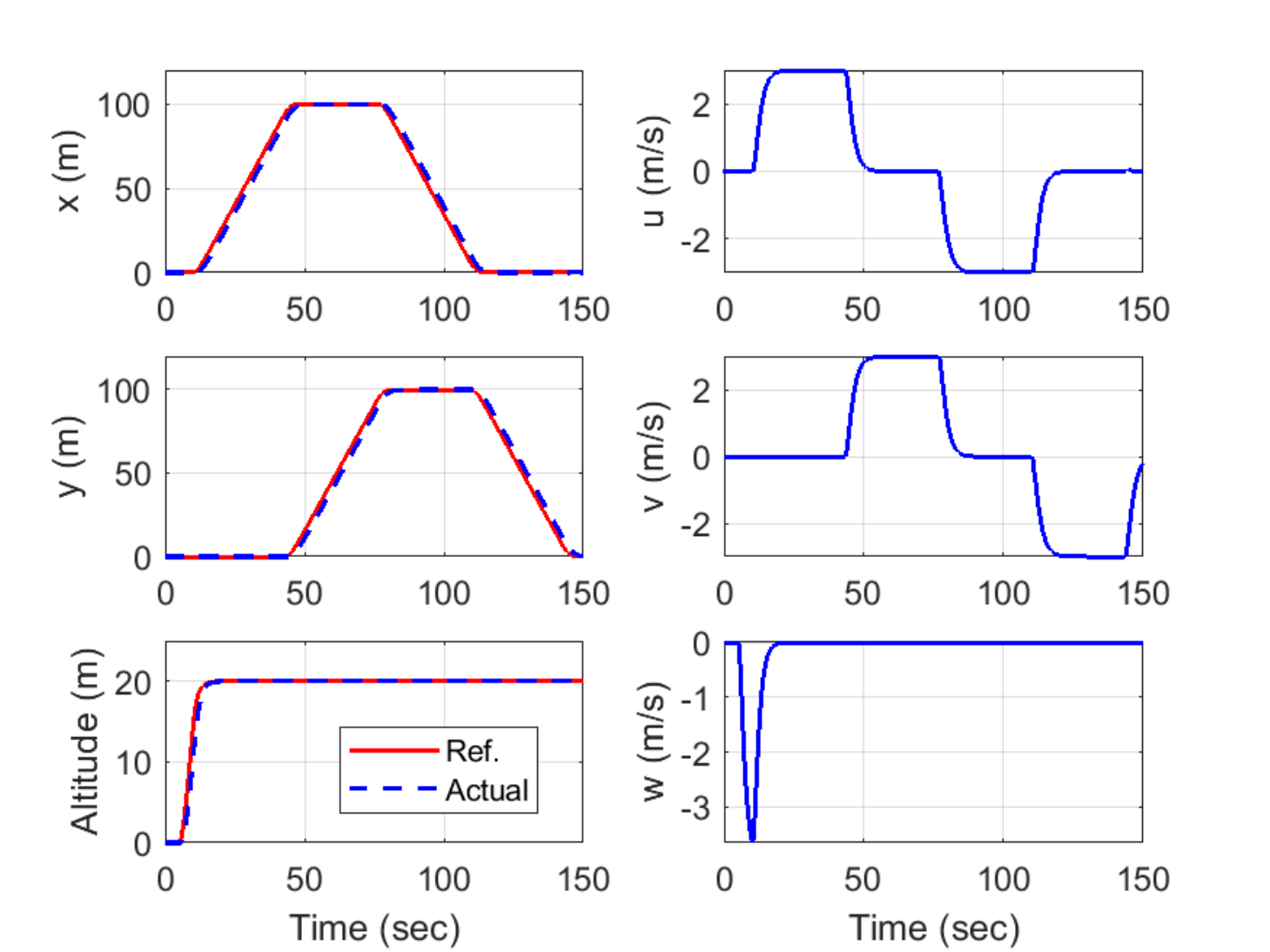}
  \caption{Uncontrollable Faults: Position \& Velocity}\label{fig:FTUCPosVel}
\end{figure}

\begin{figure}
  \centering
  \includegraphics[width=0.9\linewidth]{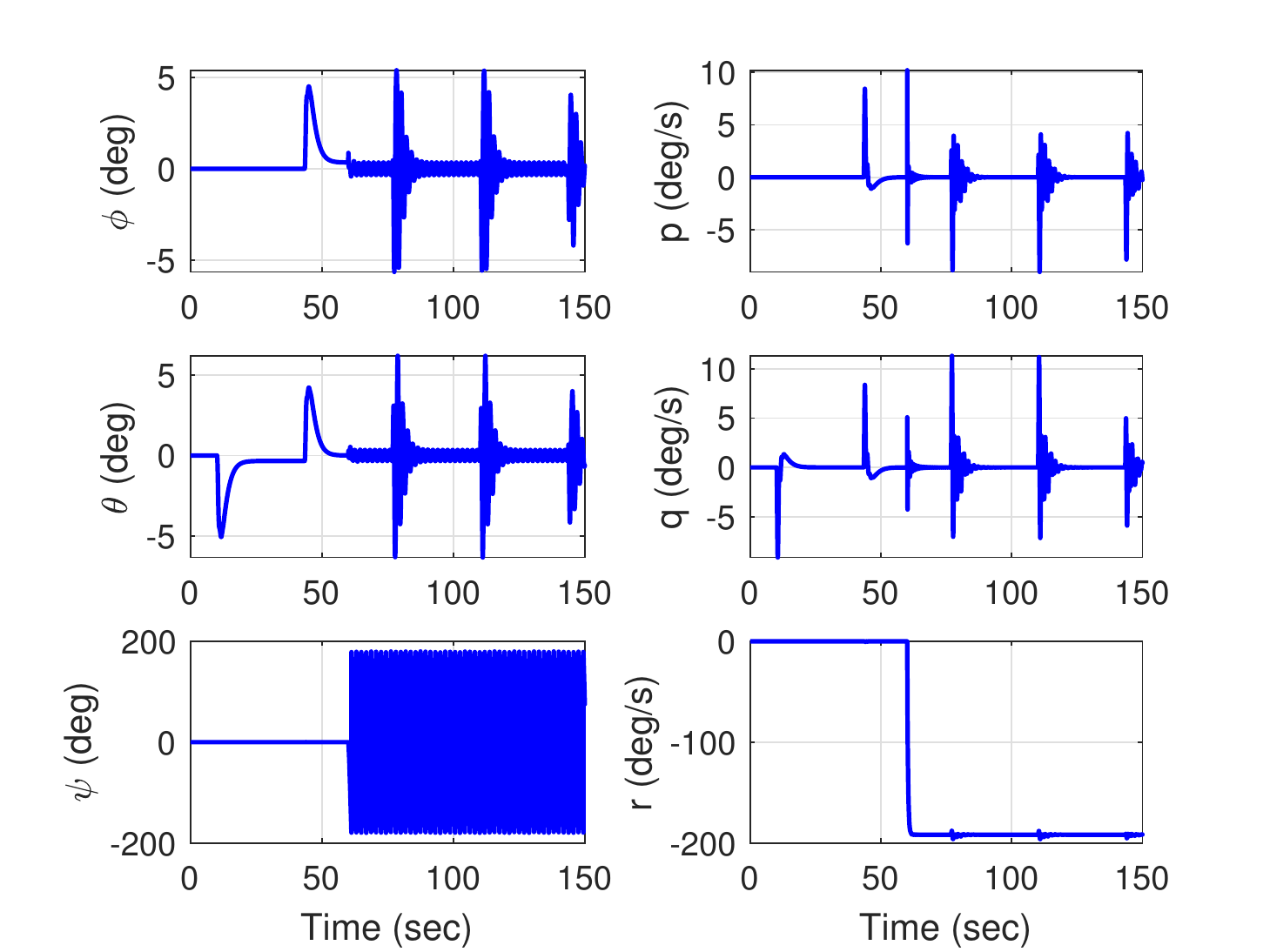}
  \caption{Uncontrollable Faults: Attitude \& Rates}\label{fig:FTUCAttitude}
\end{figure}

Fig. \ref{fig:FTUCPosVel} shows quite accurate position tracking and velocity response. Fig. \ref{fig:FTUCAttitude} shows damped but oscillatory roll and pitch behaviour, but oscillations are no more than $\pm5\,\mathrm{deg}$. However, yaw angle, being uncontrolled, increases without bound, at a constant rate of about $-200\,\mathrm{deg/s}$. This value of steady yaw-rate is purely determined by multicopter dynamics ($\kappa_{R}$) and cannot be manipulated by the controllers, when yaw channel is left uncontrolled. Results also show damped but oscillatory roll-rate and pitch-rate response within about $\pm10\,\mathrm{deg/s}$. 

\section{Conclusion}
In this work, a reconfigurable fault-tolerant control scheme for a generalized class of multicopters, namely co-planar multicopters, was presented. Initially, the controllability of multicopters in different fault scenarios was analyzed using ACAI \cite{Du2015}, and then for uncontrollable faults, the notion of reduced controllability is introduced. It was shown that even in some uncontrollable failures, it is possible to control a reduced set of states. This allows to recover and safely land multicopter even in the occurrence of uncontrollable failures, provided that reduced controllability is present. To tolerate these failures, a reconfiguration control scheme based on constrained control allocation and a robust nonlinear dynamic inversion was developed. The designed control scheme was tested and verified using high fidelity nonlinear simulations. Different controllable and uncontrollable rotor faults were injected in simulation at different flight conditions and results were presented, which shows accurate position tracking in both scenarios.

In the future, this reconfiguration control scheme can be extended to non co-planar multicopters, where it is possible, at least in some configurations, to control all six degrees of freedom independently.


\bibliography{FTC_References}
\end{document}